\documentstyle[preprint,aps]{revtex}
\begin{document}
\preprint{UM-P-94/63, RCHEP-94/17}
\draft
\title{The search for the decay of Z boson into two gammas as a
test of Bose statistics}
\author{A.Yu.Ignatiev\cite{byline1},
G.C.Joshi\cite{byline2}}
\address{Research Centre for High Energy Physics, School of
Physics, University of Melbourne, Parkville, 3052, Victoria,
Australia}
\author{M.Matsuda\cite{byline3}}
\address{Department of Physics and Astronomy, AICHI University of Education,
Igaya, Kariya, Aichi 448, Japan}
\date{June 11, 1994}
\maketitle
\begin{abstract}
We suggest that Bose statistics for photons can be tested by
looking for decays of spin-1 bosons into two photons. The
experimental upper limit on the decay $Z \rightarrow \gamma
\gamma $ is used to establish for the first time the
quantitative measure of the validity of Bose symmetry for
photons.
\end{abstract}
\pacs{}
The Standard Model of particle interactions has so far been
remarkably consistent with all the experimental data. That
adds importance to various ways of looking for possible new
physics not described by the Standard Model. Many such
possibilities have been considered before; most of them
concentrate on some kind of extension of the Standard Model
(by adding new particles, symmetries or interactions) and
then obtain from experiment various bounds on those
extensions. Here, we would like to discuss how recent
results in Z measurements can give limits on a more radical
departure from standard physics: the possible small
violation of Bose statistics.

The problem of small deviations from Fermi or Bose
statistics, initially an exotic issue raised in the early 70's,
has grown over the last decade into an elaborate area of
research leading to new ingenious experiments as well as
striking connections with modern mathematical physics
[1-34]. Yet most effort, especially in the experimental
field, was actually concentrated on discussing small
violation of the Pauli exclusion principle rather than
violation of Bose statistics. Many dedicated experiments
have been performed to give strong bounds on the violation
of the Pauli principle. By contrast, nothing similar has
been done or suggested with respect to possible deviations
from Bose statistics.

For example, there do not exist any
direct experimental bounds on Bose symmetry violation in
elementary processes (rather than statistical ensembles)
involving photons, gluons or gauge bosons. Two discussions
of experimental bounds on Bose statistics violation
which are known to us are 1) the study of the decay $K_L
\rightarrow 2 \pi^0$ in Ref. \cite{gm} from the point of
view of Bose symmetry (rather than CP) violation and 2) the
analysis of the upper limit on laser intensities implied by
the small deviations from Bose statistics \cite{fivel} (the
latter work however has been criticised in Ref.
\cite{green}). In this paper we would like to start filling
the gap by deriving the limits on Bose symmetry violations
in the system of two photons.

Before going to our main subject, a few words about the
spin-statistics theorem are in order. We would like to
remind the reader that the spin-statistics theorem proved
rigorously in the axiomatic field theory {\em does not\/}
forbid {\em small\/} violations of normal statistics. What
this theorem {\em does \/} forbid is, so to speak, {\em
``large"\/}
(or better say ``100 \%") violations of normal statistics.
More precisely, the theorem says that spin 0 fields cannot
be quantized according to Fermi (i.e., with anticommutators)
while spin 1/2 fields cannot be quantized according to Bose
(i.e., with commutators). Thus the theorem leaves open the
question whether {\em small \/} violations of statistics
exist or not.

There is a clear reason why the system of two photons is
especially interesting in testing the degree with which Bose
symmetry is exact. It has been known since the early 50's, due
to the works of Landau \cite{landau} and Yang \cite{yang}
that a pair of  photons cannot be in a state with total
angular momentum equal to unity. Therefore, the decay of any
spin-1  boson into two photons is absolutely forbidden.
Later, Nishijima \cite{nish} suggested a more direct way to
see the theorem. It is his method that we will use to
analyse the consequences of possible Bose symmetry violation
for the two-photon system.

Of all (neutral) spin 1 bosons it is natural to concentrate
on the heaviest one- the Z-boson- because one can expect
that any violations of Bose symmetry, if any, would be
better manifested  at higher energy scales.
The method is to write down the most general form of the
decay amplitude of the spin-1 particle into two photons and
then apply the conditions of gauge invariance and Bose
symmetry to that amplitude. If both conditions are applied,
the resulting amplitude is exactly zero. We are going to
show that if we impose the condition of gauge invariance but
do not require the Bose symmetry, the resulting amplitude is
not zero. This left-over amplitude depends on only one
parameter which is natural to call ``the Bose symmetry
violating parameter". We then obtain the two-gamma decay
rate of Z-boson and compare it to the experimentally known
upper bound on the branching ratio of $Z \rightarrow \gamma
\gamma $ \cite{lep}. In this way we are able for the first
time to obtain a direct bound on Bose symmetry violation for
photons.

Now, a few remarks about the relation of our method to the
existing models of small Bose symmetry violation. The most
successful model is ``the quon model" [27-34]. Quons are
particles described by the commutation relations of the
form:

\begin{equation}
a_k a_l^{\dagger}-qa_l^{\dagger}a_k = \delta_{kl},
\end{equation}
where both  quon creation operator, $a_l^{\dagger}$, and quon
annihilation operator, $a_k$, are involved. (A particular
choice $q=0$ corresponds to the case of the so-called infinite
statistics). Note that there are no commutation relations
involving only creation operators or only annihilation
operators. Moreover, such commutation relations are not
needed for calculation of matrix elements. That means, that
the above commutation relations, together with the usual
vacuum definition, $a|0 \rangle =0$,
form a perfect basis for quon quantum mechanics. If one goes
then to quon quantum field theory, then it was shown that
such theory has to be non-local, but the full details of
such theory are to be developed yet [27-34]. Because of
this, we do not try at this stage to relate our
phenomenological model of Bose symmetry violation to the
quon model. Neither do we attempt to connect our parameter
of Bose symmetry violation, see below, to the q parameter.
These problems will be considered elsewhere.

Let us turn now to our main purpose: the construction of $Z
\rightarrow \gamma \gamma $ decay amplitude. We require that
this amplitude satisfies all the standard conditions, such
as relativistic invariance and gauge invariance, but we do
not require this amplitude to be symmetric under the
exchange of  photon ends.

The most general Lorentz invariant form of the amplitude $S$
of the decay $Z \rightarrow \gamma \gamma $ is:

\begin{equation}
S(k_1, k_2, \epsilon_1, \epsilon_2)=c_{\lambda \mu \nu}(k_1,
k_2) \epsilon_0^{\lambda} \epsilon_1^{\mu} \epsilon_2^{\nu},
\end{equation}
where $k_1$ and $k_2$ are photon momenta, $ \epsilon_1$ and
$\epsilon_2$ are photon polarization vectors, $\epsilon_0$ is
Z-boson polarization vector.

Next, the condition of Lorentz invariance applied to
$c_{\lambda \mu \nu}$ leaves us with 16 possible structures
made out of the momenta $k_1$ and $k_2$ and tensors $g_{\mu
\nu}$ and $\epsilon_{\mu \nu \alpha \beta}$. But recall that
the polarization vectors must satisfy the conditions

\begin{equation}
\epsilon_1^{\mu} k_{1 \mu}=0, \; \epsilon_2^{\nu}
k_{2 \nu}=0, \; \epsilon_0^{\lambda} (k_{1 \lambda}+k_{2
\lambda})=0.
\end{equation}
Hence terms in $c_{\lambda \mu \nu}$ proportional to $k_{1
\mu}$, $k_{2 \nu}$ and $ k_{1 \lambda} + k_{2 \lambda}$  do
not contribute to $S$ and can therefore be ignored.

Thus we are left with the next most general form of
$c_{\lambda \mu \nu}$:

\begin{eqnarray}
c_{\lambda \mu \nu}=&&a_1 \epsilon_{\lambda \mu \nu \alpha}
k_{1 \alpha} + a_2 \epsilon_{\lambda \mu \nu \alpha} k_{2
\alpha}+ b_1 g_{\lambda \mu} k_{1 \nu}\nonumber\\&& + b_2 g_{\lambda \nu}
k_{2 \mu} + g g_{\mu \nu} (k_{1 \lambda}-k_{2 \lambda}) + h
(k_{1 \lambda}-k_{2 \lambda}) k_{1 \nu} k_{2 \mu}.
\end{eqnarray}

Now, the condition of the electromagnetic gauge invariance
reads

\begin{equation}
c_{\lambda \mu \nu} k_1^{\mu} \epsilon_0^{\lambda}
\epsilon_2^{\nu}=0, \label{a}
\end{equation}

\begin{equation}
c_{\lambda \mu \nu} k_2^{\nu} \epsilon_0^{\lambda}
\epsilon_1^{\mu}=0, \label{b}
\end{equation}

\begin{equation}
c_{\lambda \mu \nu} k_1^{\mu} k_2^{\nu}
\epsilon_0^{\lambda}=0. \label{b1}
\end{equation}

Going to the rest frame of Z-boson,  it can be shown that
the necessary and sufficient condition for Eq.~(\ref{a}) and
Eq.~(\ref{b}) to hold, are, correspondingly, $a_1=0$, $b_2=0$
and $a_2=0$, $b_1=0$.

After putting $ a_1=a_2=b_1=b_2=0 $ we impose the condition
Eq.~(\ref{b1}) and obtain

\begin{equation}
h=-{2g \over M_Z^2}.
\end{equation}

Therefore the most general form of the amplitude $c_{\lambda
\mu \nu}$ reduces to

\begin{equation}
c_{\lambda \mu \nu}= g(k_{1} - k_{2})_{\lambda} (g_{\mu
\nu} -{ 2 k_{1 \nu} k_{2 \mu} \over M_Z^2}). \label{c}
\end{equation}
In principle, $g$ could depend on some scalar products of
the momenta, but in our case, since all the particles are on
mass shell, we have $k_1 k_2= M_{Z}^2/2$ (and, of course,
$k_{1}^{2}=k_{2}^{2}=0$ ), so that $g$ is a pure number.
Note that the above amplitude automatically satisfies the
condition $(k_1+k_2)_{\lambda} c_{\lambda \mu \nu}=0$.

We see that this amplitude, as expected, violates Bose
symmetry because

\begin{equation}
c_{\lambda \mu \nu}(k_1, k_2)=-c_{\lambda \nu \mu}(k_2,
k_1),
\end{equation}
whereas Bose symmetry requires

\begin{equation}
c_{\lambda \mu \nu}(k_1, k_2)= +c_{\lambda \nu \mu}(k_2,
k_1).
\end{equation}
Thus the parameter $g$ can be interpreted as the parameters
of Bose statistics violation.

Now, calculating the width of the decay $Z \rightarrow
\gamma \gamma $ with the help of the amplitude Eq.~(\ref{c})
we obtain
\begin{equation}
\Gamma = {1 \over 16 \pi M_Z} |S|^2 = { M_Z\over 16 \pi} g^2.
\end{equation}
Experimentally, it has recently been measured at LEP
\cite{lep} that

\begin{equation}
BR(Z \rightarrow \gamma \gamma) < 1.4 \times 10^{-4}.
\end{equation}
Therefore

\begin{equation}
{\Gamma(Z \rightarrow \gamma \gamma) \over \Gamma_{tot}(Z)}={g^2 M_Z \over
16 \pi \Gamma_{tot}(Z)} < 1.4 \times 10^{-4}, \;
(\Gamma_{tot}(Z) \simeq 2.5 GeV).
\end{equation}
Thus, finally, we can obtain our upper bound on the Bose
violating parameter

\begin{equation}
g < 10^{-2}.
\end{equation}

The same analysis can be made for other spin-1 bosons, too, but
since they
are lighter than $Z$, one can expect that the effect of Bose
symmetry violation, if it exists at all, would be more strongly
suppressed than for the case of $Z$; that is why we do not go
into details of that.

It is possible to carry out a similar analysis for the case
of two gluons, too, but it would be much harder to get any
experimental constraints in this case.

To conclude, based on the experimental upper limit on the
decay $Z \rightarrow \gamma \gamma$ we have obtained  the
upper bound on the possible small violation of the Bose
symmetry for the system of two photons.

The authors are grateful to T.Kieu, J.-P.Ma, B.McKellar, R.Volkas
and S.Tovey for stimulating discussions. This work was
supported in part by the Australian Research Council.

\end{document}